\documentclass[runningheads]{llncs}

\usepackage{graphicx}
\usepackage{mathtools}
\usepackage{amsmath}
\usepackage{amsfonts}
\usepackage[basic]{complexity}
\usepackage{caption}
\usepackage{subcaption}
\usepackage{cite}
\usepackage{hyperref}
\usepackage[capitalise]{cleveref}

\renewcommand{\orcidID}[1]{\href{https://orcid.org/#1}{\includegraphics[scale=.03]{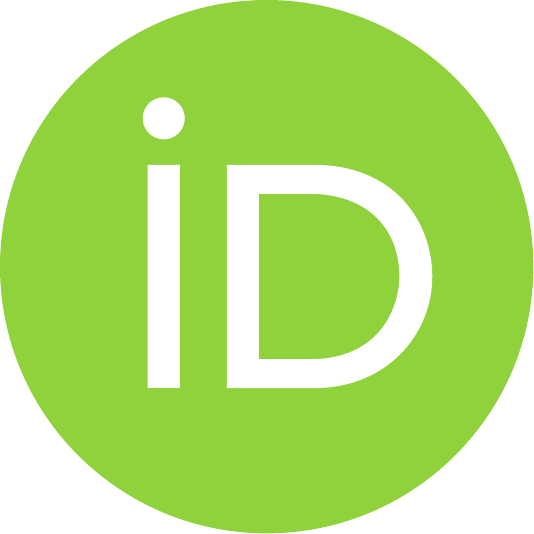}}} 

\DeclareMathOperator{\Span}{span}
\DeclareMathOperator{\Pos}{pos}

\begin{document}
\title{
Computing Hive Plots:\\ A Combinatorial Framework}
\author{Martin Nöllenburg\orcidID{0000-0003-0454-3937} \and Markus Wallinger\orcidID{0000-0002-2191-4413}}
\institute{Algorithms and Complexity Group, TU Wien \\ 
\email{\{noellenburg, mwallinger\}@ac.tuwien.ac.at}}

\maketitle              %
\begin{abstract}
Hive plots are a graph visualization style placing vertices on a set of radial axes emanating from a common center and drawing edges as smooth curves connecting their respective endpoints. In previous work on hive plots, assignment to an axis and vertex positions on each axis were determined based on selected vertex attributes 
and the order of axes was prespecified. 
Here, we present a new framework focusing on combinatorial aspects of these drawings to extend the original hive plot idea
and optimize visual properties such as the total edge length and the number of edge crossings in the resulting hive plots.  
Our framework comprises three steps: (1) partition the vertices into multiple groups, each corresponding to an axis of the hive plot; (2) optimize the cyclic axis order to bring more strongly connected groups near each other; (3) optimize the vertex ordering on each axis to minimize edge crossings. 
Each of the three steps is related to a well-studied, but \NP-complete computational problem. 
We combine and adapt suitable algorithmic approaches, implement them as an instantiation of our framework and show in a case study how it can be applied in a practical setting. Furthermore, we conduct computational experiments to gain further insights regarding algorithmic choices of the framework. The code of the implementation and a prototype web application can be found on \href{https://osf.io/6zqx9/}{OSF}\footnote{\href{https://osf.io/6zqx9/}{https://osf.io/6zqx9/ (10.17605/OSF.IO/6ZQX9)}}.

\end{abstract}
\section{Introduction}

\begin{figure}[t]
    \centering
    \includegraphics[width=0.52\textwidth]{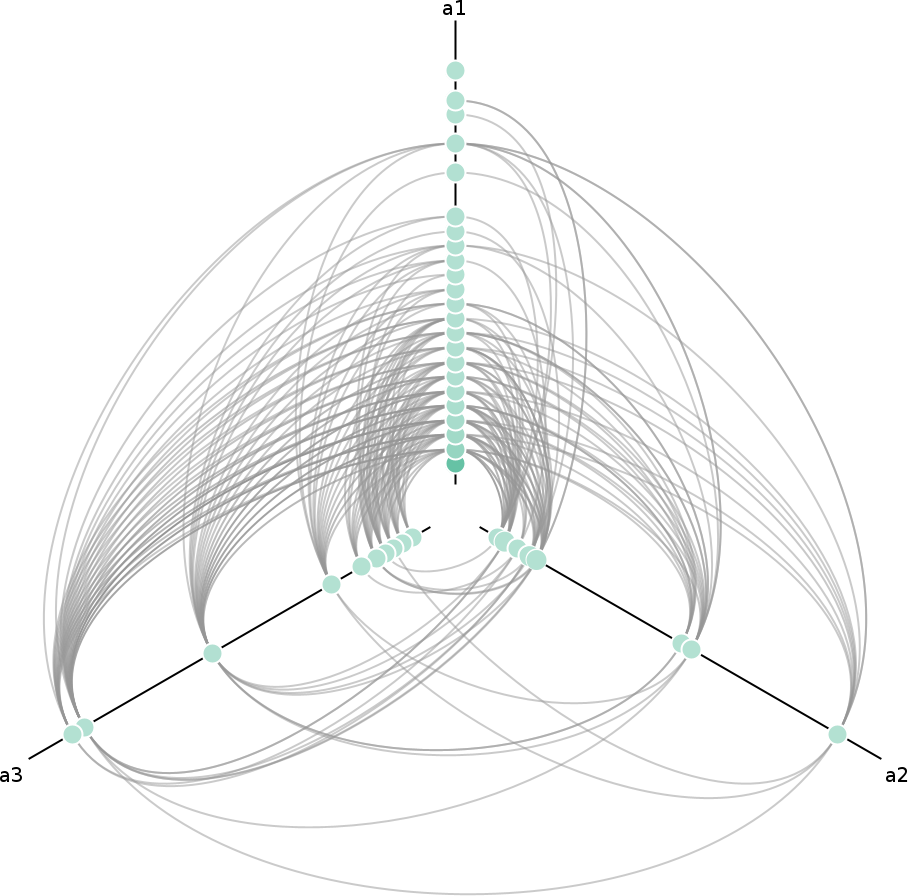}
    \caption{A hive plot created with jhive~\cite{krzywinski_hive_2012}. Vertices are mapped to axis according to their degree. The position on each axis is determined by vertex attributes such as degree (axis $a_1$) , vertex betweenness (axis $a_2$), and reachability (axis $a_3$).}
    \label{fig:original_hive}
\end{figure}

Hive plots~\cite{krzywinski_hive_2012} are a visualization style for network data, where vertices of a graph are mapped to positions on a predefined number of radially emanating axes. 
Mapping and positioning is usually done based on vertex attributes and not with the intention to optimize layout aesthetics. Due to this strict, rule-based definition, hive plots are a deterministic network visualization style; see \cref{fig:original_hive} for an example. 
Similarly to parallel coordinate plots~\cite{wegman1990hyperdimensional}, the idea behind hive plots is to quantitatively understand and compare network structures, a task that can quickly get very difficult with force-based layouts due to the 'hairball' effect for large and dense graphs and their often unpredictable behaviour when optimizing for conflicting aesthetic criteria. 

Usually, edges are drawn as Bézier curves connecting their respective end points while being restricted to three axes to avoid the problem of routing longer edges around axes; this is considered beneficial for visual clarity. 
In case of edges between vertices on the same axis, the axis and its associated vertices are cloned and positioned closely such that edges are either drawn twice (symmetrically) or once (asymmetrically).
The latter case reduces visual complexity but increases ambiguity as an edge is only explicitly connected to one copy of each vertex; see \cref{fig:hiveplot} for a sketch of  the different concepts. 

Multiple hive plots can also be arranged in a matrix, called a hive panel~\cite{krzywinski_hive_2012}, where columns and rows represent different axis mapping functions. 
Differential hive plots visualize networks changing over time~\cite{krzywinski_differential_2017}. 
Since their inception a decade ago, hive plots have been utilized in various applications and use-cases, e.g., cyber security~\cite{guarino_towards_2020}, machine learning of visual patterns~\cite{rivas_machine_2019}, life sciences~\cite{pils_cyclin_2014}, biological data~\cite{zoppi_mibiomics_2021}, or sports data~\cite{perin_soccerstories_2013}. 
Although various use-cases exist, hive plots have not yet been investigated from a formal graph drawing perspective.

This is a rather unexpected observation, especially, as hive plots have some inherent properties that make them an interesting layout style. For example, by placing vertices on axes the layout is predictable and has usually a good aspect ratio. Similarly, edges can be routed in a predictable manner. Thus, edges overlapping with unrelated vertices is not an issue in hive plot layouts and increases the overall faithfulness of the drawing. Furthermore, it is also relatively straight forward to position labels and avoid label-edge and label-vertex overlaps. Lastly, edges between vertices on the same axis can be hidden or shown on demand, thus, reducing unnecessary information and decreasing the cognitive load. 

\paragraph{Contributions and Related Work.} In this paper, we present a formal model of hive plots and identify their associated computational optimization problems from a combinatorial point of view. 
Based on this model, we investigate several unused degrees of freedom that can be utilized for optimizing hive plot layouts for arbitrary undirected graphs. 

First, in our investigation we take a new angle on assigning vertices to axes. 
Basically, the idea is to partition the graph into some number $k$ of densely connected clusters, where each cluster is assigned to exactly one axis. 
In terms of visual design this allows us to show or hide intra-cluster edges on demand and focus on representing the sparse connectivity between clusters. 
We find such clusters by applying techniques from the area of community detection in networks~\cite{fortunato_community_2010}. Even though a similar assignment strategy is presented in the original hive plot publication~\cite{krzywinski_hive_2012}, the focus there is on visually clustering vertices according to their community membership and assigning vertex clusters to segments on subdivided axis.  

Second, we are free to assign any cyclic order over the $k$ different axes. 
Here we optimize the total length of inter-axis edges by placing axes with many edges between them close to each other. This is essentially the circular arrangement problem. In the circular arrangement problem vertices are positioned evenly spaced on a circle such that the total weighted length of edges is minimized. 
Finding the minimum circular arrangement of undirected and directed graphs is \NP-complete~\cite{liberatore_circular_2004,goos_minimum_2004-1}. 
However, a polynomial-time $O(\log n)$-approximation for undirected graphs exists~\cite{liberatore_circular_2004}. 
Similarly, the problem of minimizing the crossings in a circular arrangement of a graph is \NP-complete~\cite{masuda1987np}. 
The concept of circular arrangements has been applied to circular drawings~\cite{kaufmann_improved_2007} where a subset of edges is drawn outside of the circle to reduce edge crossings.

Lastly, once the order of axes is fixed we want to minimize the number of inter-axis edge crossings. 
Here, the problem is similar to multi-layer crossing minimization which has been studied in the context of the Sugiyama framework~\cite{sugiyama_methods_1981, Tamassia_hgdv} for hierarchical level drawings of directed graphs. 
In this type of drawing vertices are assigned to horizontal layers with edges either drawn in upward or downward direction. In case of cycles in the graph, some edges need to be reversed in the drawing. 
Cyclic level drawings have already been mentioned by Sugiyama et al. as an alternativenach ablauf to reversing edges, and they have been thoroughly investigated in more recent years~\cite{bachmaier_CyclicLevelPlanarity_2008, hutchison_cyclic_2009, kaufmann_improved_2007, hutchison_global_2010}.

Crossing minimization in cyclic level drawings and layered drawings is repeatedly performing a 2-layer crossing minimization step. The 2-layer crossing minimization problem is already \NP-hard~\cite{eades_drawing_1994}, even if one layer is fixed. Heuristics~\cite{eades1994edge,kratochviyl_using_1999} have been proposed and adapted for cyclic level drawings~\cite{hutchison_global_2010}. 
We adapt the barycenter algorithm~\cite{eades1994edge} to efficiently reduce the number of crossings while adding novel constraints to force long edges to not cross over axes.
Adding constraints to 2-layer crossing minimization heuristics has been applied previously, e.g., for fixing the relative positions of a subset of vertex pairs~\cite{Forster04}.

We combined and implemented the three above mentioned problems into a 3-step framework.
Finally, we show in a case study how hive plots generated by our framework can be applied in a practical context of co-authorship networks and conduct a small-scale computational experiment on the crossing minimization aspect of our framework.

\section{Formal Model}

\begin{figure}[t]
    \centering
    \includegraphics[width=0.82\textwidth]{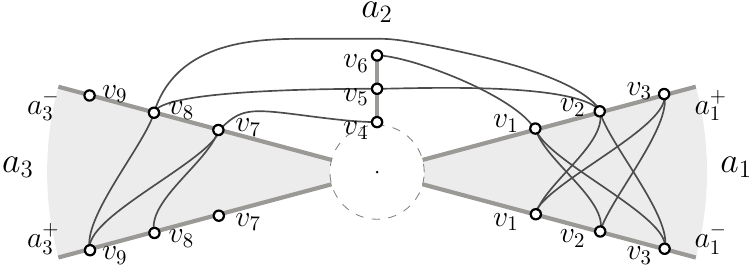}
    \caption{Schematized hive plot with three axes showing different concepts. Axis $a_2$ is collapsed. Axis $a_1$ and $a_3$ are expanded with edges in $a_1$ being drawn symmetrically. A long edge between $v_2$ and $v_8$ is bypassing $a_2$.}
    \label{fig:hiveplot}
\end{figure}

A \emph{hive plot layout} $H(G) = (A, \alpha, \phi, \Pi)$ of a graph $G=(V,E)$ is a tuple consisting of a set $A = \{a_0, \dots, a_{k-1}\}$ of $k$ axes, a surjective function $\alpha: V \to A$ mapping vertices to axes, a bijective function $\phi: A \to \{0, ..., |A| - 1\}$ representing a cyclic ordering of axes and a set $\Pi = \{\pi_0, \dots, \pi_{|A| - 1}\}$ of orderings over the vertices assigned to each axis. 
Each vertex is assigned to exactly one axis $a_i \in A$ imposing a disjoint grouping $V_i := \alpha^{-1}(a_i)$ such that $V_i \cap V_j = \emptyset$ for each $i \ne j$ with $i,j \in \{0,\dots,|A-1|\}$. 
Each $\pi_i$ is a bijective function $\pi_i: V_i \to \{0, \dots, |V_i| - 1\}$.

We use the shorthand notation $\phi(u) = \phi(\alpha(u))$ whenever the order of the axis $\alpha(u)$ of a vertex $u$ is needed. 
The \emph{span} of two axes $a_i, a_j$ or two vertices $u,v$ is defined as $\Span(a_i,a_j) = \min \{\phi(a_i) - \phi(a_j) \pmod{k}, \phi(a_j) - \phi(a_i) \pmod{k}\}$ or $\Span(u,v) = \Span(\alpha(u), \alpha(v))$.
Based on the span we can classify edges into three different categories. 
An edge $e = (u,v)$ is called \emph{proper} if $\Span(u,v) = 1$. 
Otherwise, an edge is considered \emph{long} if $\Span(u,v) > 1$ or \emph{intra-axis} if $\Span(u,v) = 0$. \emph{Inter-axis} edges are all edges that are either proper or long.
A long edge $(u,v)$ can be subdivided and replaced by $\Span(u,v) - 1$ dummy vertices assigned to the appropriate axes between $\alpha(u)$ and $\alpha(v)$.
A long edge in a hive plot layout needs to \emph{bypass} axes to connect source and target vertices without creating axis-edge overlaps. Combinatorially this can be realized by enforcing dummy vertices to appear at certain positions in each axis order. In our model, a hive plot layout can have up to $g$ gaps per axis, see \cref{fig:synthetic_example} for examples. If $g=1$ then all dummy vertices have to be at the end of each order. If $g=2$ then dummy vertices have to be at either the beginning or end of each order. In cases where $g > 2$ all dummy vertices form a partition into up to $g$ groups, where they must appear consecutively within each group.

To consider intra-axis edges during optimization an adaption is necessary. 
Basically, all axes and their associated vertices are duplicated such that for each axis $a_i$ there are two copies $a_i^+$ and $a_i^-$, respectively, and vertex sets $V_i^+$ and $V_i^-$; see \cref{fig:hiveplot}. 
The vertex order on duplicate axes remains the same, i.e.,  $\pi_i^+ = \pi_i^- = \pi_i$.

We consider two inter-axis edges $(u,v)$ and $(x,y)$ to be \emph{crossing} if $u,x \in V_i$ and $v,y \in V_{j}$ such that $\pi_i(u) < \pi_i(x)$ and $\pi_j(y) < \pi_j(v)$. 
Similarly, if the end points of two long edges $(u,v)$ and $(x,y)$ are on four different axes such that $\phi(u) < \phi(x) < \phi(v) < \phi(y) \pmod{k}$ or on three different axes such that, w.l.o.g., $\phi(u) = \phi(x) = i$, $\pi_i(u) < \pi_i(x)$, and $\phi(x) < \phi(y) < \phi(v) \pmod{k}$ a crossing is unavoidable. 
The \emph{neighborhood} of a vertex $u$ is defined as $N(u) = \{v \mid (u,v) \in E, \Span(u,v) = 1\}$.

\section{Framework for Computing Hive Plots}

Next we present our framework for creating a hive plot layout $H(G) = (A, \alpha, \phi, \Pi)$ of an undirected simple graph $G=(V,E)$. 
The framework itself is modeled as a pipeline consisting of three stages. In stage (1) we partition the vertices into multiple groups each corresponding to an axis of the hive plot. Next, we (2) optimize the cyclic axis order to bring strongly connected groups near each other. Finally, we (3) optimize the vertex ordering on each axis to minimize edge crossings. Furthermore, edge crossing minimization is performed under the constraint that long edges need to be routed through gaps in the axis. 

\subsection{Vertex Partitioning}

In the first stage we partition the vertex set $V$ into subsets $\{V_0, \dots, V_{k-1}\}$ such that each subset maps to exactly one axis $a_i$ in the hive plot.
The core idea is that the subsets of the partition represent dense induced subgraphs. 
In our implementation we use three different strategies to compute a partition. 
First, if we consider the paramter $k$ as an additional input we use the Clauset-Newman-Moore greedy modularity maximization~\cite{clauset_finding_2004} to compute a partition of size $k$.
Second, if $k$ is not specified we apply the Louvain~\cite{blondel_fast_2008} community detection algorithm instead. 
Here, the size of the partition is determined by how many communities are detected. 
Lastly, this step of the framework is not necessary if a partition is given in the input. 
Note, any other algorithm to partition the graph into meaningful groups can be used. 

\subsection{Axis Ordering}

The second stage orders the axes such that the total \emph{span} of edges is minimized. 
Our approach assumes that edges are always drawn along the shortest path around the circle between endpoints. 
Basically, we want to maximize the number of proper edges while minimizing the number and length of long edges. 
We do not consider the individual position of vertices on their respective axes, but rather look at the aggregated edges incident to the subsets of the axis partition. 

Let $w_{ij}$ be the number of edges between $V_i$ and $V_j$ for $i < j$. The cost function of an axis order $\phi$ is defined as follows:

\[
    \textrm{cost}(\phi) = \sum_{i=0}^{k - 1} \sum_{j=i+1}^{k - 1} w_{i j} \; \Span(a_i, a_j) 
\]

We can afford using an exact brute-force approach for instances with $k \le 8$  to minimize $\phi$, as it takes less than 0.5s on our reference machine (see \cref{sec:experiments}); otherwise we use simulated annealing. 

\subsection{Crossing Minimization}

In the third stage of our framework we are concerned with minimizing edge crossings under the assumption that assignment to axes and the cyclic axis order are already fixed. On each axis, the vertices are initially in random order.
Here, we employ a two-step approach, where first crossings of long edges and then intra-axis edge crossings are minimized.
Additionally, we assume that a global parameter $g \ge 1$, which represents the maximal number of gaps per axis, is given. If $g = 1$ we assume that edges are routed on the outside. In case of $g = 2$ we assume that gaps are on the outside and inside of each axis. If $g > 2$ we evenly distribute the gaps along each axis. 

First, we process all long edges to turn them into sequences of proper edges. 
Each edge $e = (u,v)$ with $\Span(u,v) > 1$ is subdivided by inserting $\Span(u,v) - 1$ dummy vertices assigned to the appropriate axes between $\alpha(u)$ and $\alpha(v)$. 
Isolated vertices which are not an endpoint of at least one long edge can be ignored in the first step of the crossing minimization.

Next, we use the barycenter heuristic~\cite{Tamassia_hgdv} for crossing minimization. 
Our approach works by iterating several times in clockwise or counter-clockwise order over all axes while performing a layer-by-layer crossing minimization sweep.
At each iteration we process all vertices of an axis by computing a new barycenter position as follows:

\[
\Pos(u) = \frac{1}{|N(u)|}\sum_{v \in N(u)} \frac{\pi_{\alpha(v)}(v)}{|\pi_{\alpha(v)}|}.
\]

As it is necessary to avoid cases where axes are imbalanced, we normalize both axes and consider the  neighbourhood $N(u)$ of vertex $u$ in the next and previous axes. 

The reason for considering both axes is that when only considering the previous axis crossings might be introduced that are overall worse 
from the reverse direction. 

\begin{figure}[t]
    \centering
    \includegraphics[width=0.82\textwidth]{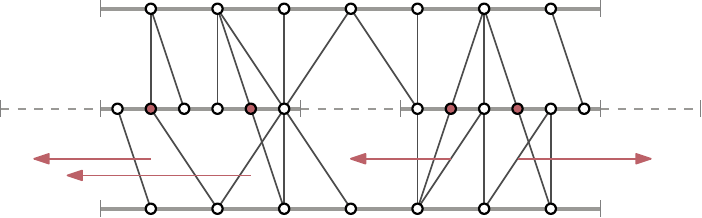}
    \caption{An axis with $g=3$ gaps. Gaps are indicated by dashed lines while axis segments are solid lines. Vertices colored red are dummy vertices. After computing new positions we move dummy vertices into either a gap to the left or right. The side is determined by counting the crossings and greedily picking the better option. The order of dummy vertices remains unchanged after the procedure.}
    \label{fig:gap_algo}
\end{figure}

Once barycenter positions are calculated we sort all vertices $v \in V_i$ of axis $a_i$ by their positions $\Pos(v)$. 
Now, to consider gaps we have to apply a case distinction on $g$.
If $g = 1$ we simply want dummy vertices on the outside to route the long edges around axes.
We constrain the sorting algorithm to put all normal vertices before all dummy vertices. 
For $g > 1$ we perform the following procedure. 
We create a list of $g$ empty lists representing the gaps and a list of $g-1$ empty lists representing the segments of the axis between gaps.
We initialize a counter $j = 0$ that represents the index of the current list. 
Next, we iteratively process vertices according to the previously computed order and distinguish between normal and dummy vertices.
Whenever we encounter a normal vertex we append it to the list of axis segments at position $j$. If the list contains more than $\frac{|V_i|}{g - 1}$ vertices we increase $j$ by one. Here, $|V_i|$ represents the number of normal vertices on axis $a_i$.
The main idea behind this is that normal vertices are evenly assigned to axis segments to increase symmetry.
If we encounter a dummy vertex we have to decide if we assign it to the gap to the left or the right of the current axis segment. 
Here, the decision can be made by looking at all vertices in the same axis segment that are to the left and compute the crossings if we put the dummy vertex in the gap to the left. Similarly, we repeat the process for the right-hand side and choose the gap which induces less crossings. Thus, appending the dummy vertex in the list at index $j$ or $j + 1$.
\Cref{fig:gap_algo} illustrates how the gap assignment works.
Finally, we assemble a new list by adding all vertices alternating between gap and axis list.
The new position of a vertex is determined by the respective index in the list.

We terminate the overall process after either no change is detected for one cycle or some iteration threshold is reached.

In the second phase of the crossing minimization we aim to further reduce intra-axis crossings by applying the barycenter heuristic. 
As the focus of the layout is on edges between different axes, we introduce the additional constraint that the relative order of vertices incident to inter-axis edges are not allowed to change in this phase any more. 
Basically, this is again a classic 2-layer crossing minimization for a vertex subset, however both layers have the same order. 
Moreover, we apply the same procedure as described above to constrain dummy vertices to gap positions.
We process each axis individually and terminate processing of an axis once no change is detected or the iteration threshold is reached.

\section{Implementation and Hive Plot Rendering}
\label{sec:rendering}

\begin{figure}[th]
     \centering
     \begin{subfigure}[b]{0.48\textwidth}
         \centering
         \includegraphics[width=\textwidth]{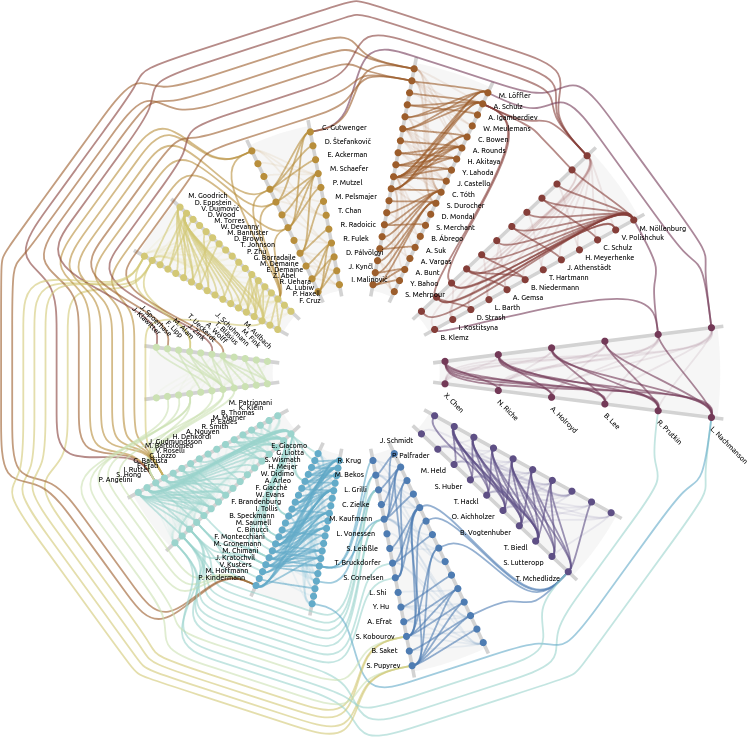}
         \caption{collapsed axes}
     \end{subfigure}
     \hfill
     \begin{subfigure}[b]{0.48\textwidth}
         \centering
         \includegraphics[width=\textwidth]{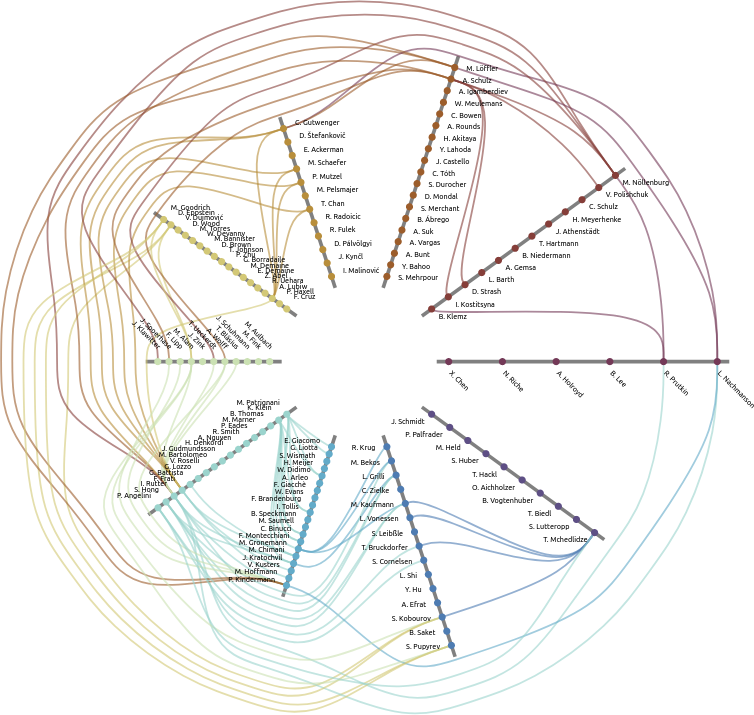}
         \caption{expanded axes}
     \end{subfigure}
     \caption{A different co-author graph from the citation dataset. The graph has 140 vertices and 397 edges with 47 being proper and 37 being long. Here, simulated annealing was used and the total computation time of the layout was approximately 51ms.}
     \label{fig:large1}
\end{figure}

In this section we will briefly explain the design decisions regarding the visualization; see \cref{fig:CS} or \cref{fig:large1} for examples.
The implementation code and a prototype web application can be found on \href{https://osf.io/6zqx9/}{OSF}.
Axes are drawn as straight lines emanating from a common center with equal angular distribution. 
Axes can be expanded or collapsed on demand. 

When an axis is expanded the background is colored in a light grey color with low opacity. 
When expanded, the available space is distributed $40:60$ between intra-axis and inter-axis edges. 
Vertices are drawn as small circles and their positions on their respective axis $a_i$ are computed based on their position in $\pi_i$. 
Labels are placed in clockwise direction next to an axis horizontally. If a vertex' assigned axis differs by $\pm$ 25$^{\circ}$ to the horizontal reference direction, the label is rotated by 45$^{\circ}$.
Edges are drawn as Bezíer curves. 
For edges between neighbouring axes control points are set perpendicular to the axes. 
If long edges are routed around axes, the positions of their dummy vertices are converted to control points. 
The color of vertices is computed by mapping the angle of the assigned axis to a radial color map~\cite{crameri2020misuse}. 
For edges, we assign the color of the first endpoint in counter-clockwise direction. 
The ideas behind coloring edges are that it becomes easier to follow individual edges and that it is, for half of the vertices, immediately clear to which axis the edge connects.

\paragraph{Interactivity.}
\Cref{fig:CS_inter} shows an example how interactivity was realized in our visualization. 
First, when hovering a vertex, the vertex itself, all neighbours and incident edges are highlighted by a color contrasting the color scheme. 
Initially, each axis in the visualization is collapsed. 
By clicking on a single axis it is expanded to show more details on demand.
Furthermore, it is also possible to expand all axes with a button click.
Naturally, it is also possible to unexpand individual axis. 
Similarly, by clicking a button in the interface vertices can be scaled to represent their respective degree; see \cref{fig:CS_scaled}.
Lastly, labeling can be toggled on or off.

\section{Case Study}
\label{sec:case_study}

We evaluate our framework by a case study using the \emph{citation} dataset~\cite{devanny2017graph} from the creative contest at the 2017 Graph Drawing conference.
This dataset contains all papers published at GD from 1994 to 2015. 
We created co-authorship graphs for different years by extracting researchers from papers and connecting them with edges whenever they co-authored a paper.

\begin{figure}[!ht]
     \centering
     \begin{subfigure}[b]{\textwidth}
         \centering
         \includegraphics[width=\textwidth]{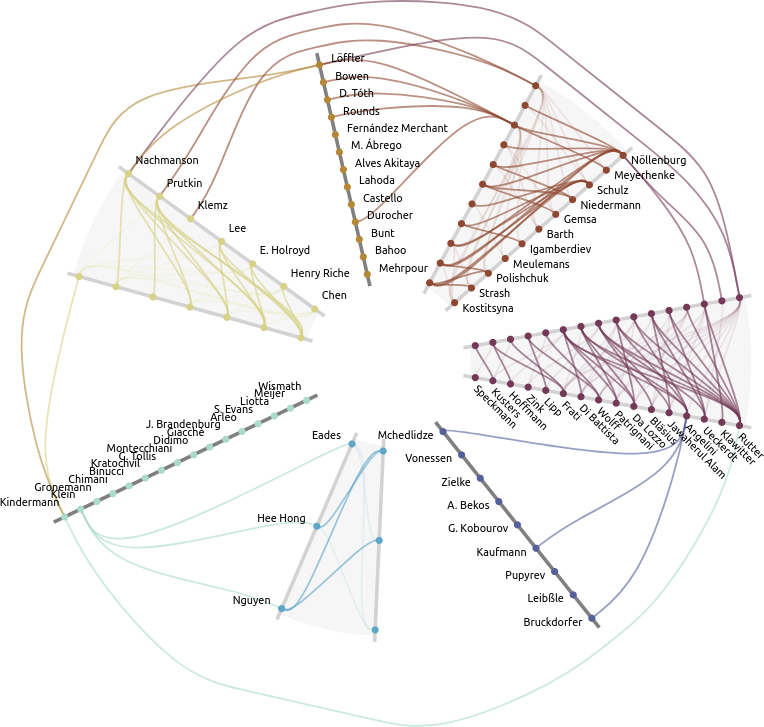}
         \caption{}
         \label{fig:CS_collapsed}
     \end{subfigure}
     \hfill
     \begin{subfigure}[b]{0.32\textwidth}
         \centering
         \includegraphics[width=\textwidth]{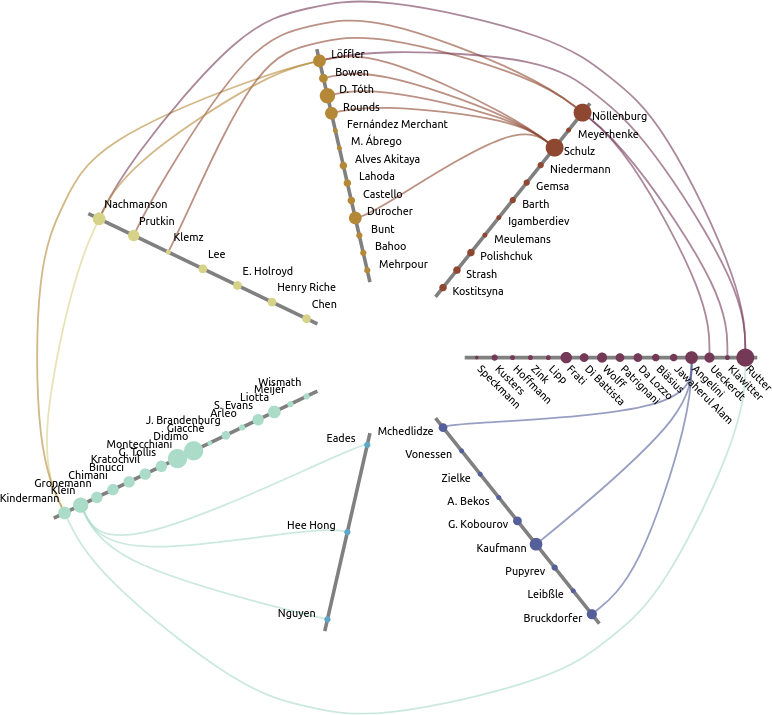}
         \caption{}
         \label{fig:CS_scaled}
     \end{subfigure}
     \begin{subfigure}[b]{0.32\textwidth}
         \centering
         \includegraphics[width=\textwidth]{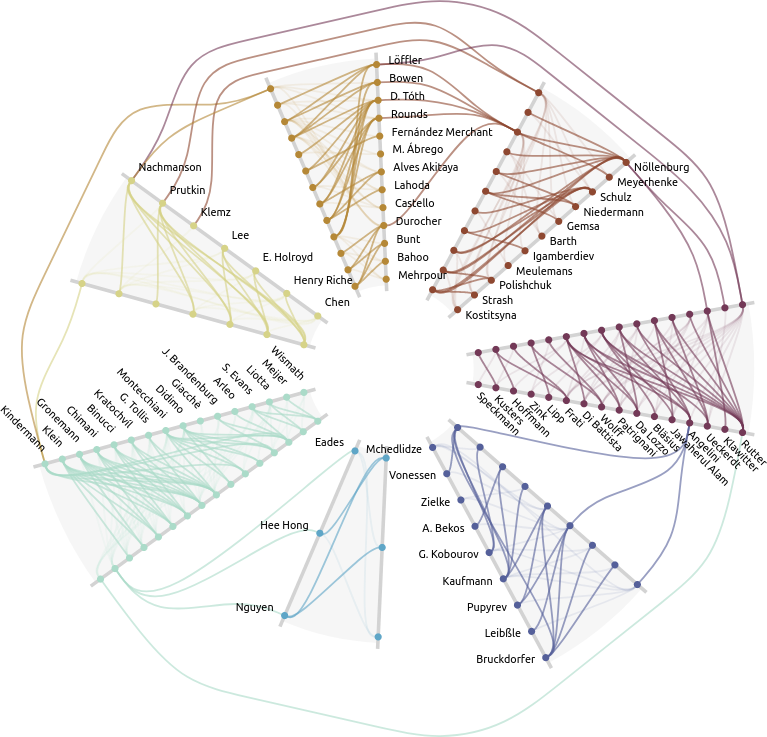}
         \caption{}
         \label{fig:CS_expanded}
     \end{subfigure}
     \begin{subfigure}[b]{0.32\textwidth}
         \centering
         \includegraphics[width=\textwidth]{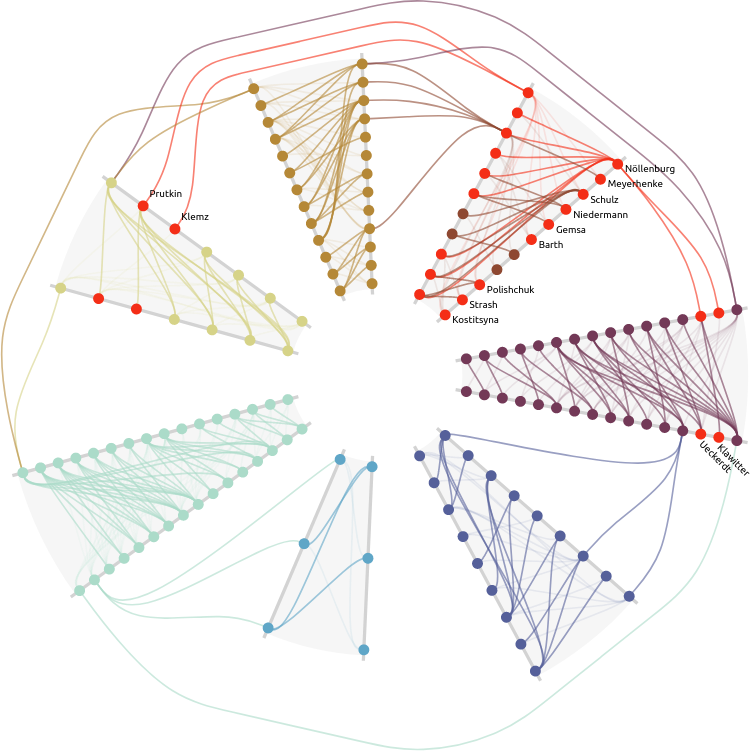}
         \caption{}
         \label{fig:CS_inter}
     \end{subfigure}
    \caption{Variations of the co-author graph of GD 2015. In (a) some axes of interest are expanded while (c) shows all axes expanded. In (b) vertices are scaled by degree and all axes are collapsed. In (d) interactive highlighting is obtained by hovering the vertex ``M. Nöllenburg''.}
    \label{fig:CS}

\end{figure}

In \cref{fig:CS} we show a hive plot of the co-author network of 2015 and three alternative renderings computed with our framework in less than 10ms. 
We used the rendering style described in \cref{sec:rendering}. We did not specify the number of axes $k$ in the input.
We specified the number of gaps as $g = 1$.
The network has a total of 75 vertices, which are partitioned into 7 groups. 
A total of 190 edges is split into 172 intra-axis edges, 12 proper edges and 6 long edges.

Authors mapped to individual axes seem to represent mainly clusters of geographic proximity of researchers' institutions or established close collaborations. 
Inter-axis edges are emphasized in our hive plots and indicate collaborations between clusters in the form of researchers bridging institutions and forming new connections, e.g., via papers originating from research visits or recent changes in affiliation.
Another possible interpretation can be seen when vertices are scaled by degree. 
Researchers with connections to other axes are often also highly connected inside their own cluster. 
This could mean that they are well connected and prolific and use existing connections to start new collaborations.

\begin{figure}[!ht]
     \centering
     \begin{subfigure}[b]{0.78\textwidth}
         \centering
         \includegraphics[width=\textwidth]{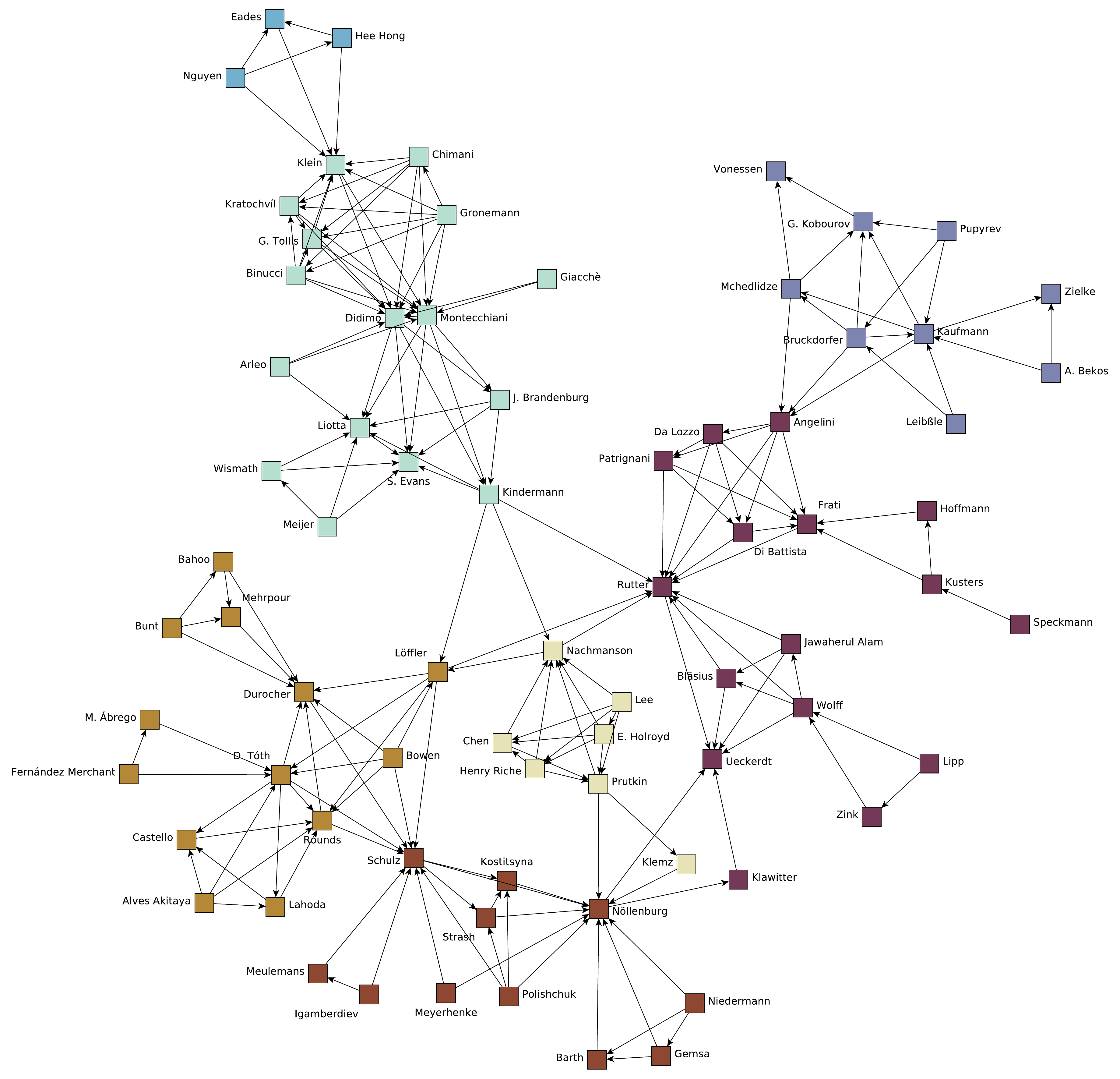}
         \caption{force-based}
         \label{fig:fb_layout}
     \end{subfigure}
    \par\bigskip
     \begin{subfigure}[b]{\textwidth}
         \centering
         \includegraphics[width=\textwidth]{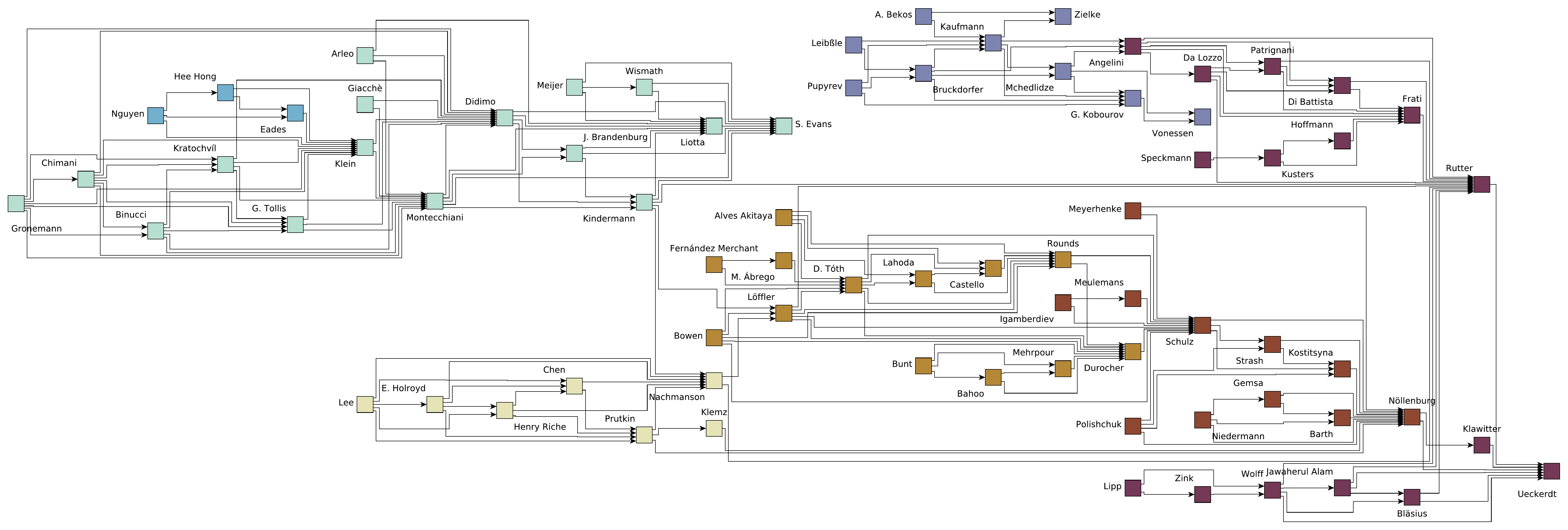}
         \caption{hierarchical}
         \label{fig:hier_layout}
     \end{subfigure}
     
     \caption{Force-based layout (a) of the co-authorship graph of Section~\ref{sec:case_study} created with yEd~\cite{yed_citation}. The smart organic layout functionality was used. The preferred edge length was set to 100 while the minimal vertex distance was set to 20. The option of avoid vertex/edge overlaps was active with a value of 0.8. Also, the labeling was optimized and the graph colored according to the partitions from the case study. (b) shows a hierarchical layout created with yEd. Similar layout optimization steps were applied.}
     \label{fig:case_study_comparision}
\end{figure}

In contrast to a force-based layout, see \cref{fig:fb_layout}, several observations can be made. While cliques are very prominent in the force-based layout the macro structure of the graph is less clear. The hive plot layout on the other hand focuses more on the macro structure of the graph with intra-axis structure only shown on demand. However, with two copies per vertex, cliques are harder to identify. Still, the hive plot layout requires no additional cue, such as color in this case, to highlight the community structure. Furthermore, in the hive plot layout individual vertices are easier to identify, labels are more uniform and edges are routed in a predictable manner which is similar to schematic diagrams. Due to the possibility of expanding axes on demand in the hive plot layout, individual communities can be easily explored without being overwhelming. While it is possible to represent communities in the force-based layout by meta vertices, it is not straight-forward to encode the relationship of each single vertex to the rest of the network. Both layouts show some label-edge overlap. Finally, the hive plot layout has a more balanced space utilization.

Furthermore, we also compared against a hierarchical layout; see \cref{fig:hier_layout}. Here, we assumed edge direction from the hive plot layout by directing edges clockwise. Naturally, the hierarchical layout emphasizes the imposed hierarchy while the communities are dispersed over the layout. Still, the communities are visible although it is questionable without the use of coloring.   This gives the layer assignment a different meaning than our approach of axis assignment. 
The orthogonal layout of edges initially simplifies the process of following an edge, but it becomes progressively more challenging with an increase in bends and crossings.
In the hive plot layout this is less of an issue, especially for edges between communities. 
Lastly, the label placement in the hierarchical layout is optimized and avoids label-edge and label-vertex overlaps. However, this optimization comes at the cost of requiring more space. In contrast, the hive plot layout has a few label-edge overlaps but utilizes space more efficiently.

\clearpage

\section{Computational Experiments} \label{sec:experiments}

We conducted a small-scale computational experiment which focused on the implications of using different numbers of gaps while minimizing edge crossings. The datasets and the evaluation code can again be found on \href{https://osf.io/6zqx9/}{OSF}.

\begin{figure}[!t]
     \centering
     \begin{subfigure}[b]{0.31\textwidth}
         \centering
         \includegraphics[width=\textwidth]{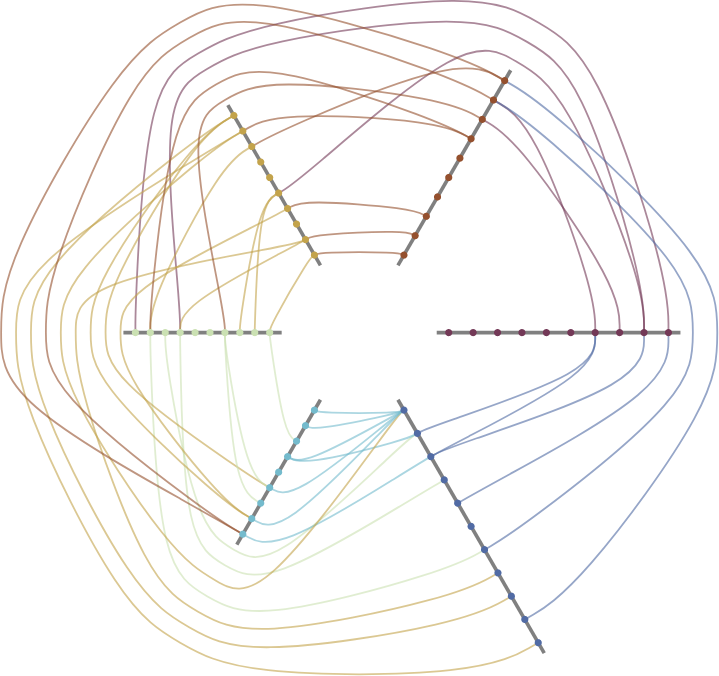}
         \caption{one gap}
     \end{subfigure}
     \hfill
     \begin{subfigure}[b]{0.31\textwidth}
         \centering
         \includegraphics[width=\textwidth]{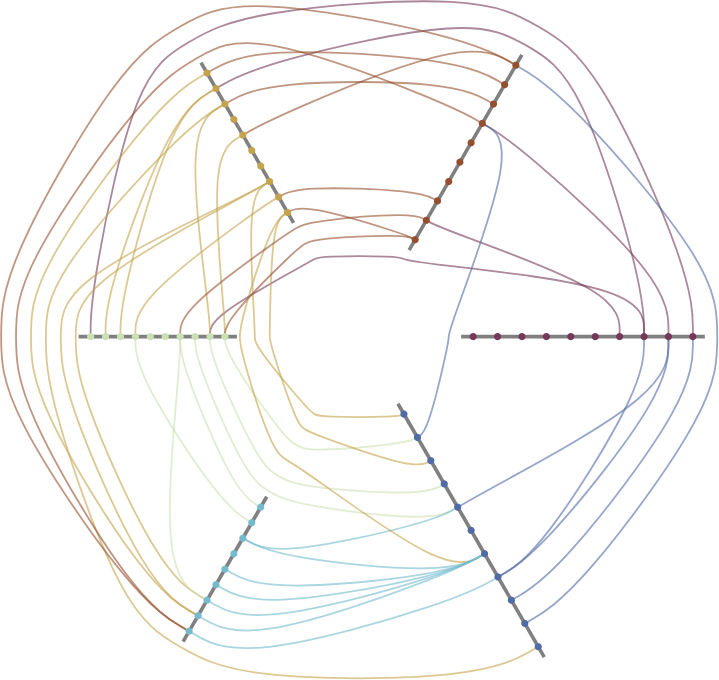}
         \caption{two gaps}
     \end{subfigure}
     \hfill
     \begin{subfigure}[b]{0.31\textwidth}
         \centering
         \includegraphics[width=\textwidth]{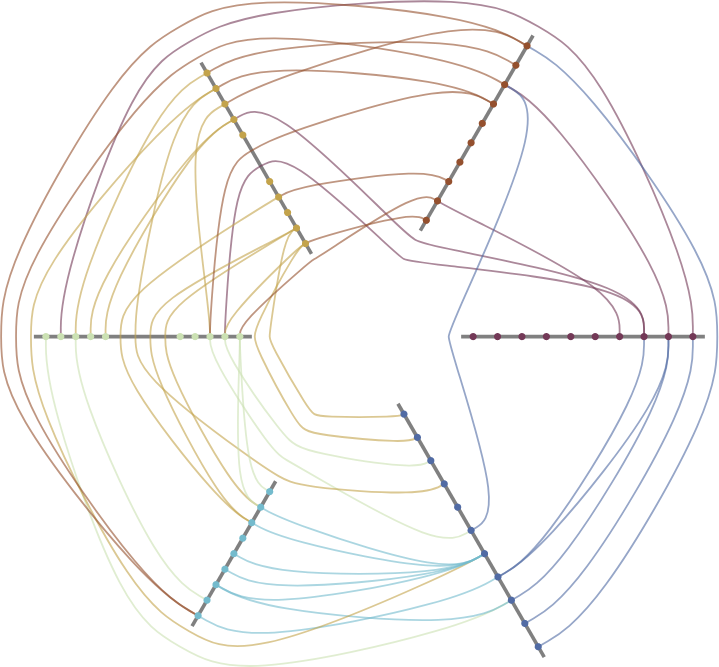}
         \caption{three gaps}
     \end{subfigure}
     \caption{One example graph from the synthetic datasets used in the computational experiment. (a) shows one gap on the outside, (b) shows an additional second gap on the inside and (c) allows for an additional third gap in the middle of each axis.}
     \label{fig:synthetic_example}
\end{figure}

\paragraph{Dataset and Setup.} First, we created a dataset of synthetic graphs that tries to capture variations of input sizes similar to the use-case of visualizing communities in a small to medium size graph. Examples can be seen in \cref{fig:synthetic_example}. To generate the graphs we used a random partition graph~\cite{fortunato_community_2010} implementation of networkX.
We varied the number $n$ of vertices from $60$ to $510$ with a step size of $30$ with a fixed number of six partitions $\{V_1,\dots,V_6\}$ where each partition was of size $|V_i| = \frac{n}{6}$. 
Furthermore, we had to specify the probability of edges between vertices of the same and different partitions. 
For edges between vertices of the same partition we set the probability to $p_{in} = \frac{6}{|V_i|}$ which gives an expected average of connecting a vertex to six other vertices in the same partition. 
For edges between vertices of different partitions we set $p_{out} = \frac{2}{n - |V_i|}$ which connects a vertex on expected average to two other vertices outside of its own partition.
For each step size we generated five graphs. In contrast to using real-world data, the above procedure gives us a predictable size of the graph that has a decent number of long and proper edges.  

The implementation to compute a hive plot is written in Python 3.11. All experiments where run on a machine with Ubuntu 22.04, 16GiB of RAM and an Intel i7-9700 CPU with 3.00Hz.

\paragraph{Experiment.} In our experiment we evaluated the effect of varying the number of gaps in the input. We computed a hive plot layout for all graphs in the dataset with fixed $k=6$ but varied $g \in \{1,2,3\}$. 

\begin{figure}[!t]
     \centering
     \begin{subfigure}[b]{0.49\textwidth}
         \centering
         \includegraphics[width=\textwidth]{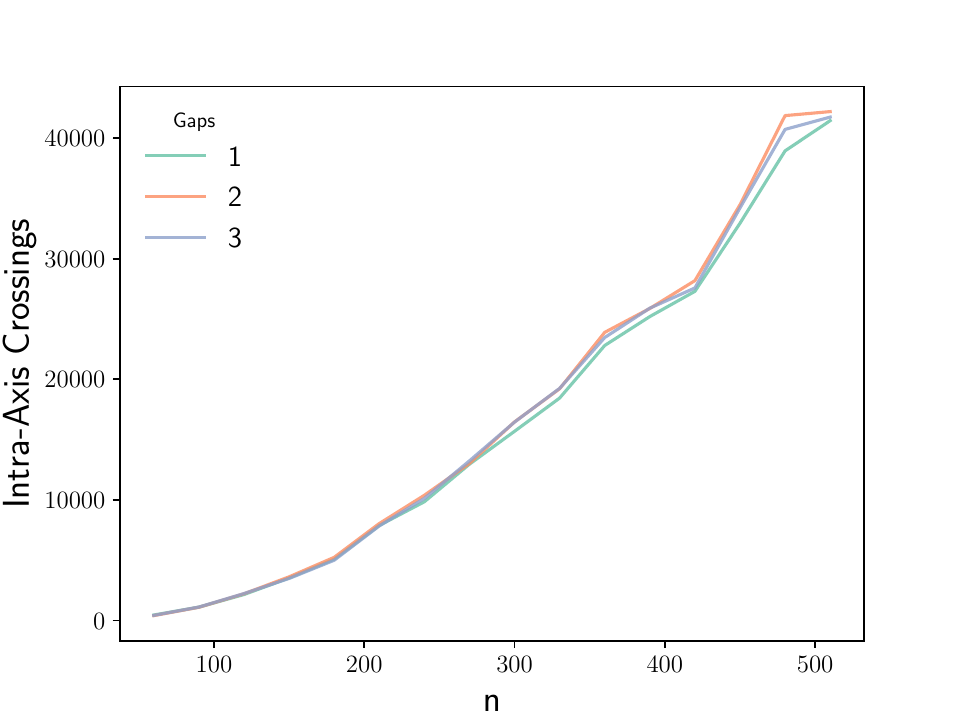}
         \caption{intra-axis crossings}
         \label{fig:exp_gaps_1}
     \end{subfigure}
     \hfill
     \begin{subfigure}[b]{0.49\textwidth}
         \centering
         \includegraphics[width=\textwidth]{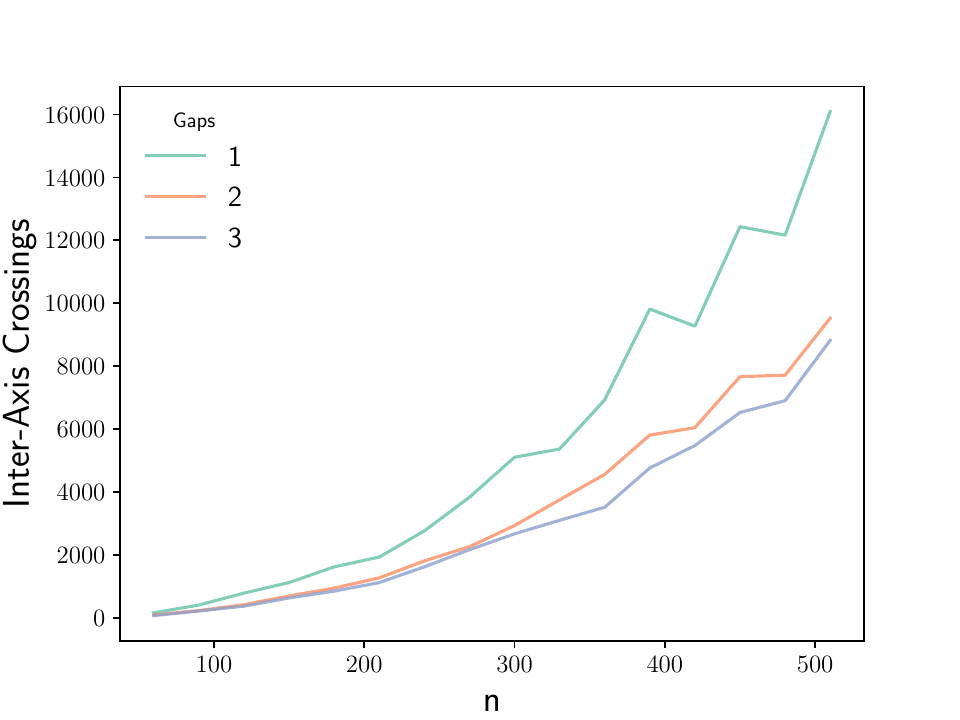}
         \caption{inter-axis crossings}
         \label{fig:exp_gaps_2}
     \end{subfigure}

     \begin{subfigure}[b]{0.49\textwidth}
         \centering
         \includegraphics[width=\textwidth]{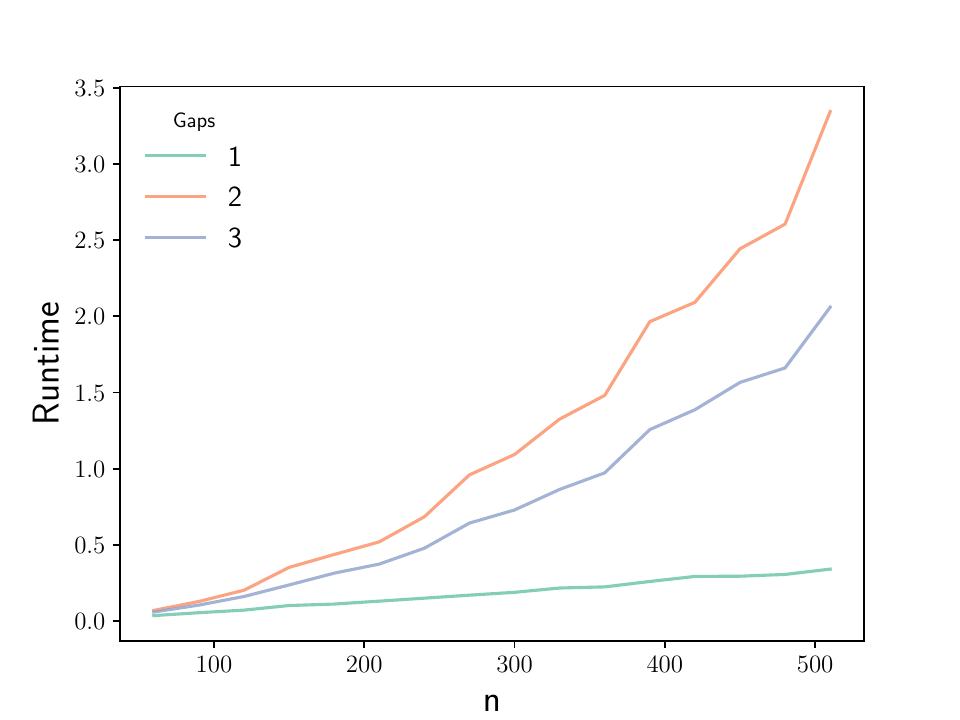}
         \caption{runtime}
         \label{fig:exp_gaps_3}
     \end{subfigure}
    \caption{Number of crossings and runtime for the synthetic datasets. The x-axis shows an increase in number of vertices which correlates with the number of edges. The y-axis in (a) and (b) show the total number of crossings. In (c) the runtime for our hive plot framework is shown.}
    \label{fig:exp_gaps}
\end{figure}

We counted the number of crossings of intra-axis edges and the number of crossings of inter-axis edges. We did not consider crossings of inter-axis edges with intra-axis edges as this can only be observed if $g \ge 3$. The resulting plots can be seen in \cref{fig:exp_gaps_1} and \cref{fig:exp_gaps_2}. Interestingly, the number of intra-axis crossings does not substantially differ between the three variants. On the other hand, using two or three gaps drastically decreases the number of inter-axis crossings. While a difference between two and three gaps is visible in the plots, it is not as strong as between one and two, or one and three, gaps. Thus, it is questionable whether sacrificing the zero crossings between intra-axis and long inter-axis edges one gets with only two gaps is worth the slight reduction in inter-axis edge crossings.  

The runtime plot can be seen in \cref{fig:exp_gaps_3}. The runtime for one gap stays consistently below 0.4s while the runtime for two and three gaps increases much steeper. Interestingly, the runtime for three gaps is lower than for two gaps. This can be explained by moving dummy vertices to gap positions. For two gaps we have to inspect all vertices in the axis to the left and right while for three gaps we have to inspect at most half of the vertices as we can stop once we inspect a vertex that is assigned to a different segment of an axis.

\section{Discussion and Conclusion}

We have introduced a framework for drawing hive plots. 
Our edge routing guarantees that vertices are never occluded by edges, which is generally not the case in frequently used algorithms such as force-based layouts. 
The focus of our approach lies on showing inter-axis connections, i.e., the long edges, which reduces the visual complexity, but at the same time emphasizes the weak ties in networks, which are highly important in network analysis~\cite{g-swt-73}. 
Nonetheless, interactivity can be used to show intra-axis edges on demand and thus give a more detailed view on the dense parts of the network. 
Since the aspect ratio of hive plots is fixed, the layout can easily be integrated into a multi-view visual analytics system. 

Obviously, there are also some limitations to our approach. 
The original hive plots~\cite{krzywinski_hive_2012} are deterministic renderings of networks based on vertex attributes, whereas the current implementation of our approach partially uses non-deterministic algorithms, e.g., for the clustering step, which may lead to different hive plots for the same data or data with small changes. 
In the clustering step there are other approaches that were not considered in our prototype implementation that could potentially improve the layout.
Even for an optimal axis order the presence of many long edges decreases the readability quickly. 
The angular resolution for more than 8-12 axes becomes too small to precisely show connectivity details, especially for vertices closer to the origin. 
Therefore, our framework has limited visual scalability and is recommended mostly for
small to medium graphs with less than 500 vertices and no more than 8--12 clusters. 
However, it is possible to hide some visual complexity by collapsing individual axes.
Lastly, the currently implemented crossing minimization heuristic is based on the barycenter algorithm. Here, more sophisticated approaches, such as sifting~\cite{MatuszewskiSM99} could potentially be adapted to incorporate gaps.

In terms of future work, several questions arise. 
While our choice of algorithms in the framework was mostly guided by best-practice from existing literature, a more thorough investigation regarding exact solutions or bounds on typical quality criteria could lead to interesting insights. 
Also, instead of using a multi-stage framework it could be possible to optimize for multiple criteria simultaneously. 
Similarly, we only looked at aggregates of axes and the cyclic length when optimizing order, which does not consider the actual Euclidean length of edges. 
Potentially, we can improve scalability in the number of axes if we arrange them on an ellipse to increase space between axes. 
From a human-computer interaction perspective it would be interesting to see how our hive plot framework compares to other layouts for visualizing small to medium sized graphs in a formal human-subject study. 
Finally, adding more interactivity and integrating our framework as an alternative view for data exploration into a visual analytics platform could provide additional insights.

\bibliographystyle{splncs04}
\bibliography{bibliography}

\end{document}